\IfFileExists{IEEEtran.cls}
  {\documentclass[conference]{IEEEtran}}
  {\documentclass[10pt,twocolumn]{article}\usepackage[margin=0.75in]{geometry}}
\usepackage[utf8]{inputenc}
\usepackage[T1]{fontenc}
\usepackage{lmodern}
\usepackage{amsmath,amssymb,amsthm}
\usepackage{graphicx}
\usepackage{booktabs}
\usepackage{xcolor}
\usepackage{tikz}
\usetikzlibrary{positioning,arrows.meta,fit,backgrounds}
\usepackage{url}
\usepackage[hidelinks]{hyperref}
\newcommand{\code}[1]{\texttt{#1}}
\newtheorem{proposition}{Proposition}
\newtheorem{definition}{Definition}

\title{Samyama-SDG: A Declarative GraphQL \code{@count} and Regex/Ontology Synthetic Data Compiler with $O(1)$-Memory Referential Integrity and Causal Ground Truth}
\author{%
  \IEEEauthorblockN{Madhulatha Mandarapu}
  \IEEEauthorblockA{Samyama.ai\\madhulatha@samyama.ai}
  \and
  \IEEEauthorblockN{Sandeep Kunkunuru}
  \IEEEauthorblockA{Samyama.ai\\sandeep@samyama.ai}
}

\begin{document}
\maketitle

\begin{abstract}
Synthetic data generators face a divide between declarative relational mockets and domain causal simulators. Declarative generators specify entities, regexes, and foreign keys via schemas, but buffer parent primary-key pools in RAM ($O(N)$ memory) and lack ontology bindings or causal ground truth. Conversely, domain simulators emit realistic streams and ground-truth labels, but require thousands of lines of procedural code per domain.

We present \textbf{Samyama-SDG} (\code{sdg-rs}), a Rust synthetic data compiler unifying declarative schema compilation with ontology-bound causal simulation. At \textbf{Layer 1}, Samyama-SDG compiles Cardinality-Extended GraphQL schemas (\code{@count}, \code{@pk}, \code{@ontology}, \code{@regex}, \code{@dist}, \code{@fk}, \code{@after}) and an Attribute Data Dictionary into a topologically sorted DAG. Combining stateless counter-based seekable PRNGs (SplitMix64 over $(\text{seed}, \text{entity}, \text{row}, \text{field})$) with index-bijective primary keys $f_{\text{PK}}(s, E, i)$ lets child tables resolve uniform or Zipfian foreign keys and multi-hop \code{@after} constraints in $O(1)$ memory without buffering parent tables. At \textbf{Layer 2}, causal processes layer fault propagation and machine-scorable ground truth (\code{ground\_truth.jsonl}) over the same seekable PRNG and \code{samyama-ontologies} binder. Across three schemas, Samyama-SDG generates $1{,}000{,}000$ records in $128.79\text{ ms}$--$172.43\text{ ms}$ ($5.80\text{M}$--$7.76\text{M}$ rows/s) and $5{,}000{,}000$ records in $676.12\text{ ms}$--$808.56\text{ ms}$ at flat $31.4\text{ MiB}$--$35.1\text{ MiB}$ memory, with $0$ conformance errors across $1{,}150{,}000$ foreign keys and $900{,}000$ temporal constraints.
\end{abstract}

\begin{IEEEkeywords}
synthetic data generation, GraphQL schema compilation, referential integrity, counter-based PRNG, property graphs, ontology binding, causal ground truth
\end{IEEEkeywords}

\section{Introduction}\label{sec:intro}

Benchmarking relational databases, property graph engines, and machine-learning pipelines requires synthetic datasets that simultaneously satisfy structural integrity, statistical realism, vocabulary compliance, and reproducibility. Over the past three decades, synthetic data generation has evolved along three largely separate tracks:

\begin{enumerate}
\item \textbf{Declarative schema and constraint generators.} Systems following Gray et al.~\cite{gray1994quickly}, Bruno and Chaudhuri's Data Generation Language (DGL)~\cite{bruno2005flexible}, Houkj{\ae}r et al.~\cite{houkjær2006simple}, and modern developer tools allow users to declare table schemas, attribute regular expressions, and foreign-key relationships. While onboarding a new domain requires only a short schema file, existing implementations typically store generated parent primary keys in in-memory arrays so child tables can sample valid foreign keys. Consequently, memory consumption grows linearly ($O(N)$) with dataset scale, and sequential stateful random number generators couple column values: adding one attribute to a schema shifts the random stream for every subsequent attribute and table.
\item \textbf{Tabular statistical and deep generative models.} Frameworks such as the Synthetic Data Vault (SDV)~\cite{patki2016sdv} and CTGAN~\cite{xu2019ctgan} learn joint distributions from existing production tables. However, when cold-starting a greenfield application, ontology, or benchmark, no training corpus exists; furthermore, neural tabular generators struggle to guarantee $100\%$ multi-table referential integrity or exact multi-hop temporal ordering at million-row scale.
\item \textbf{Domain-specific state-machine and causal simulators.} Healthcare generators such as Synthea~\cite{walonoski2018synthea}, graph benchmarks such as LDBC SNB~\cite{erling2015ldbc}, and our Telecom NOC alarm-storm generator~\cite{mandarapu2026alarm} simulate rich domain mechanisms over time. These systems produce high-realism event streams and ground-truth labels, but require writing hundreds to thousands of lines of custom imperative code per domain.
\end{enumerate}

\paragraph{First-Principles Unification.} This work grew out of reconciling two artifacts built three years apart within our own engineering practice: an April 2023 declarative prototype that compiled Cardinality-Extended GraphQL schemas (\code{type Entity @count(N)}) and a Regex/Enum Data Dictionary into relational tables, and our 2026 ontology-bound Telecom NOC causal generator~\cite{mandarapu2026alarm} built on \code{samyama-ontologies} (\code{NORIA-O}~\cite{tailhardat2024noria}, \code{RFC 8632}~\cite{vallin2019rfc8632}, and \code{TMF642}) and Internet Topology Zoo backbones~\cite{knight2011internet}. Analyzing both from first principles revealed that declarative structural compilation and causal ground-truth simulation are not competing paradigms, but \emph{two layers of a single synthetic data compiler}:

\begin{itemize}
\item \textbf{Layer 1 (Structural \& Relational/Graph Compiler):} Compiles declarative entity-relationship topologies, attribute regexes/distributions, ontology class/enum bindings, and temporal ordering constraints into either relational CSV tables or property graph nodes and edges (\code{nodes.jsonl}, \code{edges.jsonl}).
\item \textbf{Layer 2 (Causal Process \& Ground-Truth Engine):} Executes stateful domain processes (such as network fault injection, reachability loss, and alarm storm emission) over a Layer 1 inventory graph, emitting an operational event stream (\code{alarms.stream.jsonl}) alongside a held-out causal ground-truth log (\code{ground\_truth.jsonl}).
\end{itemize}

\begin{figure*}[t]
\centering
\begin{tikzpicture}[
  inb/.style={draw, rounded corners=3pt, minimum height=0.82cm, align=center, font=\footnotesize, fill=blue!8},
  cmp/.style={draw, rounded corners=3pt, minimum height=0.88cm, align=center, font=\footnotesize, fill=gray!10},
  eng/.style={draw, rounded corners=3pt, minimum height=0.92cm, align=center, font=\footnotesize, fill=green!10},
  outb/.style={draw, rounded corners=3pt, minimum height=0.82cm, align=center, font=\footnotesize, fill=orange!12},
  arr/.style={-{Stealth[length=2.2mm]}, thick}
]
  \node[inb, text width=3.5cm] (gql) at (0, 1.1) {\textbf{GraphQL SDL Schema} \\ \code{type E @count(N) @pk} \\ \code{@fk(dist, s)} \code{@after}};
  \node[inb, text width=3.5cm] (dict) at (0, 0.0) {\textbf{Attribute Data Dictionary} \\ Regex ASTs, Distributions \\ Semantic Pools (\code{dictionary.yaml})};
  \node[inb, text width=3.5cm] (onto) at (0, -1.1) {\textbf{Ontology Builds} \\ \code{samyama-ontologies} \\ (\code{NORIA-O}, \code{RFC 8632}, \code{TMF642})};

  \node[cmp, text width=3.7cm] (dag) at (4.7, 0.55) {\textbf{Layer 1 Schema Compiler} \\ Back-Ref Pruning + Kahn DAG \\ Regex AST \& Ontology Binder};
  \node[cmp, text width=3.7cm] (pack) at (4.7, -0.85) {\textbf{Layer 2 Causal Pack} \\ Real Backbone (\code{GraphML}) \\ Fault \& Propagation Rules};

  \node[eng, text width=4.0cm] (rng) at (9.4, 0.55) {\textbf{Stateless Streaming Engine} \\ Seekable PRNG $\phi(s, E, i, A)$ \\ $O(1)$-RAM Bijective FKs $f_{\text{PK}}(s, E, j)$};
  \node[eng, text width=4.0cm] (sim) at (9.4, -0.85) {\textbf{Causal Storm Simulator} \\ Sub-Stream Isolated Stages \\ Held-Out \code{ground\_truth.jsonl}};

  \node[outb, text width=3.5cm] (out1) at (14.0, 0.65) {\textbf{Dual Structural Output} \\ Relational \code{<Entity>.csv} \\ Graph \code{nodes.jsonl}, \code{edges.jsonl}};
  \node[outb, text width=3.5cm] (out2) at (14.0, -0.75) {\textbf{Provenance \& Validation} \\ \code{binding.json}, \code{manifest.json} \\ \code{realism.json} ($0$ errors)};

  \draw[arr] (gql.east) -- (dag.west);
  \draw[arr] (dict.east) -- (dag.west);
  \draw[arr] (onto.east) -- (dag.west);
  \draw[arr] (onto.east) -- (pack.west);
  \draw[arr] (dag.east) -- (rng.west);
  \draw[arr] (pack.east) -- (sim.west);
  \draw[arr] (rng.east) -- (out1.west);
  \draw[arr] (rng.east) -- (out2.west);
  \draw[arr] (sim.east) -- (out1.west);
  \draw[arr] (sim.east) -- (out2.west);
\end{tikzpicture}
\caption{Architecture of the Samyama-SDG compiler (\code{sdg-rs}). Layer 1 compiles Cardinality-Extended GraphQL SDL schemas, an Attribute Data Dictionary, and \code{samyama-ontologies} builds into a lock-free Rayon streaming plan with $O(1)$-memory index-bijective foreign keys. Layer 2 layers causal fault propagation and ground-truth logging over the same seekable PRNG and vocabulary binder.}
\label{fig:arch}
\end{figure*}
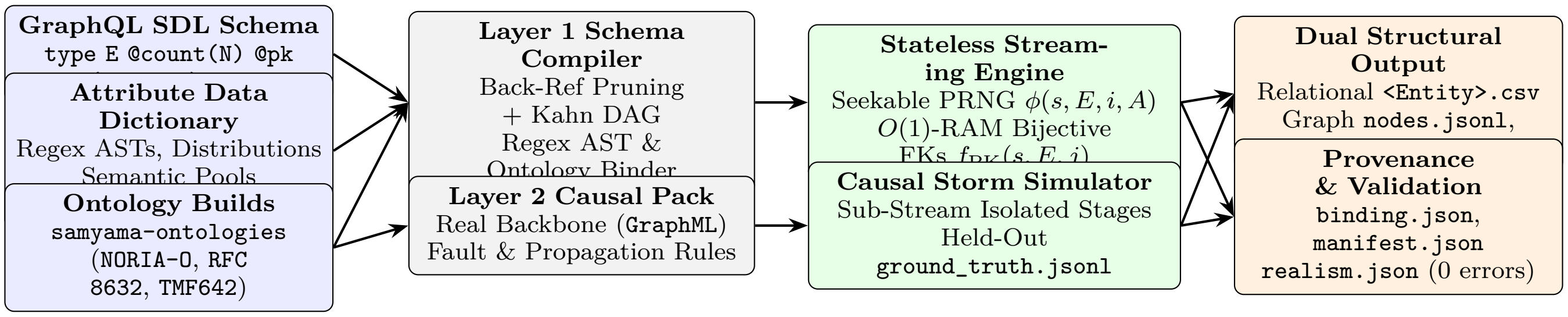

\paragraph{Contributions.} We design, implement in Rust, and evaluate \textbf{Samyama-SDG} (\code{sdg-rs}, Fig.~\ref{fig:arch}):
\begin{itemize}
\item \textbf{C1 --- Cardinality-Extended GraphQL \& Data Dictionary Compiler (\S\ref{sec:lang}):} We formalize a declarative schema language extending GraphQL SDL with \code{@count(n)}, \code{@pk}, \code{@ontology}, \code{@regex}, \code{@dist}, \code{@fk}, and \code{@after} directives, backed by a reusable Attribute Data Dictionary and compile-time \code{samyama-ontologies} verification.
\item \textbf{C2 --- Stateless Counter-Based Seekable PRNG with Schema Evolution Invariance (\S\ref{sec:rng}):} Replacing sequential stateful PRNGs with a deterministic hash-mixing function $\phi(s, E, i, A)$ over SplitMix64~\cite{steele2014fast} and Lemire's divisionless interval sampling~\cite{lemire2019fast} makes every cell an $O(1)$ pure function of its coordinates, enabling lock-free Rayon data parallelism and proving that adding or reordering schema fields leaves all other generated columns bit-for-bit identical (Proposition~\ref{prop:invariance}).
\item \textbf{C3 --- $O(1)$-Memory Index-Bijective Foreign Keys and Recursive Temporal Resolution (\S\ref{sec:fk}):} By defining primary keys $f_{\text{PK}}(s, E, i)$ and base dates $f_{\text{Date}}(s, E, i, A)$ as pure functions of the target row index $i \in [0, N_E)$, child tables sample parent indices $j \sim \mathcal{D}_{\text{FK}}([0, N_P))$ under uniform or power-law Zipfian ($\text{Zipf}(s)$) distributions and synthesize both the parent primary key and multi-hop \code{@after} dates in $O(1)$ memory without materializing parent tables in RAM (Proposition~\ref{prop:o1fk}).
\item \textbf{C4 --- Multi-Scale Evaluation \& Native Rust Causal Simulation (\S\ref{sec:eval}):} Across Healthcare ($5$ tables), E-Commerce ($4$ tables), and University ($4$ tables) schemas from $1{,}000$ to $5{,}000{,}000$ records, \code{sdg-rs} achieves up to $7.76\text{M}$ rows/s in CSV mode and $2.57\text{M}$ rows/s in dual CSV + Property Graph JSONL mode at flat $31.4\text{ MiB}$--$35.3\text{ MiB}$ CSV resident memory, while accelerating the Layer 2 Telecom NOC generator by $2.67\times$ on \code{small.yaml} and $1.99\times$ on \code{large.yaml}.
\end{itemize}

\section{Multi-Frontend Declarative Schema and Data Dictionary Compiler}\label{sec:lang}

Rather than coupling the execution engine to a single input syntax, Samyama-SDG follows an LLVM-style compiler design: three interchangeable schema frontends---\textbf{Cardinality-Extended GraphQL SDL} (\code{*.graphql}), \textbf{SQL DDL} (\code{*.sql}), and \textbf{OWL/RDFS/SKOS Turtle Ontologies} (\code{*.ttl})---compile alongside a \emph{Global Attribute Data Dictionary} (\code{dictionary.yaml}) into a single typed Intermediate Representation (\code{CompiledSchema}).

\subsection{Cardinality-Extended GraphQL SDL, SQL DDL, and OWL/Turtle Frontends}

Standard GraphQL SDL defines object types, fields, and enumerations, while SQL DDL (\code{CREATE TABLE ... REFERENCES ... CHECK (...)}) and OWL/RDFS Turtle (\code{owl:Class}, \code{owl:ObjectProperty}, \code{owl:DatatypeProperty}, \code{skos:ConceptScheme}) serve relational database engineers and ontology authors directly. Across all three frontends, compiler directives annotate types, columns, and properties:
\begin{enumerate}
\item \code{@count(n: Int!)} on \code{type}: specifies the base cardinality $N_E^{(0)}$ of entity $E$. At runtime, passing \code{--scale $\alpha$} sets $N_E = \max(1, \lfloor \alpha \cdot N_E^{(0)} \rfloor)$, while \code{--count E=K} overrides a specific entity count.
\item \code{@pk(kind: "uuid"|"ulid"|"prefixed", prefix: "...")} on \code{type}: binds the entity's primary key to a deterministic index-bijective scheme (\S\ref{sec:fk}).
\item \code{@ontology(vocab: "...", class: "...", enum: "...")} on \code{type} or \code{enum}: binds the entity label or enumeration values to a compiled vocabulary in \code{samyama-ontologies}. During compilation, the ontology binder verifies that every referenced class and enum symbol exists in the vocabulary's \code{classes.json} or \code{enums.json} and records the input file's SHA-256 digest in \code{binding.json}. Any unknown term halts compilation immediately.
\item \code{@regex(pattern: "...")} and \code{@constraint(pattern: "...", min: ..., max: ...)} on a field: compiles a regular expression into a zero-allocation syntax tree (\code{CompiledRegex}) supporting character classes (\code{[A-Z]}, \code{\textbackslash d}), literals, bounded repetitions (\code{\{m,n\}}), and alternations (\code{(a|b)}).
\item \code{@dist(kind: "...", min: ..., max: ...)} and \code{@weights(values: [...])} on a field or enum: overrides the attribute's statistical distribution (\code{uniform\_int}, \code{uniform\_float}, \code{normal}, \code{lognormal}, \code{poisson}, \code{bool}, \code{date}, or weighted categorical).
\item \code{@fk(dist: "uniform"|"zipf", s: Float, edge: "...")} on a field whose type is another entity $P$: declares a foreign-key reference to $P$ (and a corresponding directed edge label in property-graph mode) sampled either uniformly or from a power-law $\text{Zipf}(s)$ distribution.
\item \code{@after(field: "parent.attr"|"localAttr", min\_days: Int, max\_days: Int)} on a \code{Date} field: enforces causal temporal ordering such that the generated date lies in $[t_{\text{ref}} + \text{min\_days}, t_{\text{ref}} + \text{max\_days}]$.
\item \code{@derived(formula: "...")} and \code{@edge(from: "...", to: "...", type: "...")}: evaluates intra-row arithmetic/boolean formulas (e.g., \code{balance - holds}, \code{amount > 500}) in $O(1)$ space and emits attributed property-graph edges directly into \code{edges.jsonl}.
\end{enumerate}

\paragraph{Virtual Reverse Back-References and DAG Compilation.} GraphQL schemas frequently include bidirectional navigation fields, such as \code{appointments: [Appointment!]!} on \code{Patient} alongside \code{patient: Patient!} on \code{Appointment}. In a naive parser, mutual references create a dependency cycle $\text{Patient} \leftrightarrow \text{Appointment}$. Samyama-SDG's compiler classifies any list-typed entity field (\code{[Child!]!}) as a \emph{virtual reverse back-reference} materialized implicitly by the child's scalar foreign key, prunes it from the dependency graph, and orders all entities via Kahn's topological sort~\cite{kahn1962topological}. True scalar foreign-key cycles ($A \to B \to A$) are rejected at compile time with an explicit cycle diagnostic.

\subsection{Global Attribute Data Dictionary}

Rather than repeating regexes or distributions on every field, Samyama-SDG resolves an unannotated field \code{fieldName: ScalarType} through a three-tier hierarchy:
\begin{equation}
\small
\text{Rule}(E, A) = \text{Dir}(E, A) \oplus \text{ByName}(A) \oplus \text{ByScalar}(\text{type}(A))
\label{eq:lookup}
\end{equation}
where $\oplus$ denotes left-biased precedence. The built-in dictionary ships with compiled regex ASTs and curated pools for clinical, financial, and network types (\code{SSN}, \code{Email}, \code{Phone}, \code{ICD10}, \code{CPT}, \code{NDC}, \code{SKU}, \code{IPv4}, \code{Date}, \code{Time}, \code{Money}, and human/organization name pools), extensible via \code{dictionary.yaml}.

\section{Theoretical Foundation: Seekable PRNGs and $O(1)$-Memory Integrity}\label{sec:rng}

\subsection{Stateless Counter-Based Seekable PRNG}

Traditional synthetic data generators maintain a single mutable PRNG state $S_{t+1} = \text{step}(S_t)$ advanced sequentially as rows and columns are emitted. This design has two structural flaws: (1) multi-threaded generation requires either coarse locking or fragile chunk-seed handoffs, and (2) adding a single attribute to entity $E$ consumes extra random draws per row, altering every subsequent attribute and entity in the dataset.

Following counter-based PRNG principles~\cite{salmon2011parallel,steele2014fast}, Samyama-SDG eliminates mutable generator state across rows and columns. Let $s \in \mathbb{Z}_{2^{64}}$ be the master seed, $h_{\text{FNV}}(x) \in \mathbb{Z}_{2^{64}}$ be the 64-bit FNV-1a hash of identifier string $x$, and $\text{mix}_{64}: \mathbb{Z}_{2^{64}} \to \mathbb{Z}_{2^{64}}$ be the SplitMix64 bijective finalizer~\cite{steele2014fast}:
\begin{align}
z_1 &= (z \oplus (z \gg 30)) \cdot c_1 \pmod{2^{64}}, \nonumber\\
z_2 &= (z_1 \oplus (z_1 \gg 27)) \cdot c_2 \pmod{2^{64}}, \nonumber\\
\text{mix}_{64}(z) &= z_2 \oplus (z_2 \gg 31),
\label{eq:splitmix}
\end{align}
with SplitMix64 multipliers $c_1 = \text{\code{0xbf58476d1ce4e5b9}}$ and $c_2 = \text{\code{0x94d049bb133111eb}}$. For any entity $E$, row index $i \in [0, N_E)$, and attribute $A$, we construct an isolated cell PRNG state $R(s, E, i, A)$ initialized at counter $c = 0$ with stream key:
\begin{equation}
\small
\phi(s, E, i, A) = \text{mix}_{64}\!\big(\text{mix}_{64}(\text{mix}_{64}(s \oplus h_{\text{FNV}}(E)) + i) \oplus h_{\text{FNV}}(A)\big).
\label{eq:cellkey}
\end{equation}
Each draw $k \in \{0, 1, \dots\}$ within cell $(E, i, A)$ evaluates $\text{mix}_{64}(\phi(s, E, i, A) + k \cdot \gamma)$, where $\gamma = \text{\code{0x9e3779b97f4a7c15}}$ is the golden-ratio odd constant, and maps 64-bit words to bounded intervals $[0, M)$ via Lemire's divisionless 128-bit multiply-high algorithm~\cite{lemire2019fast}.

\begin{proposition}[Schema Evolution Invariance and Lock-Free Parallelism]\label{prop:invariance}
Let $\mathcal{S}$ and $\mathcal{S}'$ be two schemas that share an entity $E$ with cardinality $N_E$ and a non-dependent scalar or foreign-key attribute $A$. If $\mathcal{S}'$ is obtained from $\mathcal{S}$ by inserting, deleting, or reordering any other attributes $A' \neq A$ on $E$, or by modifying any unrelated entity $E' \neq E$, then for every seed $s$ and row index $i \in [0, N_E)$, the value emitted for $(E, i, A)$ under $\mathcal{S}'$ is bit-for-bit identical to that emitted under $\mathcal{S}$, independent of thread count or row scheduling order.
\end{proposition}
\begin{proof}
By Eq.~\eqref{eq:cellkey}, the initial state $\phi(s, E, i, A)$ depends solely on $(s, h_{\text{FNV}}(E), i, h_{\text{FNV}}(A))$. No draw from any $(E, i, A')$ with $A' \neq A$ or $(E', i', \cdot)$ with $E' \neq E$ mutates $\phi(s, E, i, A)$. Because $i$ is an explicit coordinate argument, partitioning $[0, N_E)$ across arbitrary Rayon worker threads yields the exact same sequence of cell states.
\end{proof}

\subsection{$O(1)$-Memory Index-Bijective Foreign Keys and Temporal Chains}\label{sec:fk}

To avoid buffering parent primary-key tables in memory, we make every entity's primary key a deterministic bijection of its row index.

\begin{definition}[Index-Bijective Primary Key]\label{def:pk}
For a master seed $s$, entity $E$ with cardinality $N_E$, and row index $i \in [0, N_E)$, the primary-key synthesizer $f_{\text{PK}}(s, E, i)$ is a deterministic injective map defined by the entity's \code{@pk} directive:
\begin{itemize}
\item \textbf{Prefixed (\code{kind: "prefixed"}):} $f_{\text{PK}}(s, E, i) = \text{format}(\text{\code{"\{\}-\{:08d\}"}}, \text{prefix}(E), i + 1)$.
\item \textbf{Deterministic UUIDv4 (\code{kind: "uuid"}):} Combines $w_{\text{hi}} = \text{mix}_{64}(\phi_{\text{PK}}(s, E) + 2i)$ and $w_{\text{lo}} = \text{mix}_{64}(\phi_{\text{PK}}(s, E) + 2i + 1)$ with RFC 4122 version-4 and variant-1 bitmasks.
\item \textbf{Deterministic ULID (\code{kind: "ulid"}):} Encodes a deterministic timestamp prefix and 80-bit SplitMix64 payload into Crockford Base32.
\end{itemize}
\end{definition}

When child entity $C$ at row $i_C$ evaluates a foreign-key attribute $A_{\text{FK}}$ targeting parent entity $P$ (which has known compiled cardinality $N_P$), $C$ does not look up a stored array of $P$'s keys. Instead, $C$ draws a target parent index $i_P \in [0, N_P)$ from its isolated cell stream $R_C = R(s, C, i_C, A_{\text{FK}})$:
\begin{equation}
\small
i_P = \begin{cases}
\text{UniformInt}(R_C,\, 0, N_P - 1) & \text{if uniform,} \\
\big\lfloor N_P \cdot ((N_P^{1-s_{\text{z}}} - 1) u + 1)^{\frac{1}{1-s_{\text{z}}}} \big\rfloor & \text{if } \text{Zipf}(s_{\text{z}}),
\end{cases}
\label{eq:fksample}
\end{equation}
where $u \in [0, 1)$ is a uniform float drawn from $R(s, C, i_C, A_{\text{FK}})$ and $s_{\text{z}} > 1$ is the Zipf skew exponent. The child then evaluates $f_{\text{PK}}(s, P, i_P)$ in $O(1)$ time and $O(1)$ space.

\paragraph{Recursive Cross-Entity Temporal Resolution (\code{@after}).} A subtler challenge arises when a child's date depends on a parent's date, which itself depends on a grandparent's date—for example, in \code{health\_clinic.graphql}:
\begin{equation*}
\small
\text{Patient.dob} \preceq \text{Appt.date} \preceq \text{Diag.date} \preceq \text{Claim.filedDate}.
\end{equation*}
In a naive streaming generator, when generating \code{Claim} row $i_{\text{clm}}$ referencing \code{Appointment} row $i_{\text{apt}}$, the generator does not know \code{Appointment} $i_{\text{apt}}$'s date unless all $400{,}000$ \code{Appointment} dates were buffered in RAM. Samyama-SDG solves this in $O(1)$ memory via recursive coordinate replay: to evaluate $\text{evalDate}(s, E, i, A_{\text{date}})$ for an attribute with directive \code{@after(field: "fk.parentDate", min\_days: $d_{\min}$, max\_days: $d_{\max}$)}, the engine:
\begin{enumerate}
\item Re-computes the foreign-key target index $i_P \leftarrow \text{sampleFK}(s, E, i, \text{fk})$ in $O(1)$ time using Eq.~\eqref{eq:fksample}.
\item Recursively evaluates $t_P \leftarrow \text{evalDate}(s, P, i_P, \text{parentDate})$ in $O(d)$ steps, where $d$ is the schema DAG depth (typically $d \le 4$).
\item Adds the offset $\Delta \in [d_{\min}, d_{\max}]$ drawn from $R(s, E, i, A_{\text{date}})$ to $t_P$.
\end{enumerate}

\begin{proposition}[$O(1)$-Memory Referential and Temporal Integrity]\label{prop:o1fk}
For any acyclic schema $\mathcal{S}$ of DAG depth $d$ and any scale factor $\alpha > 0$, streaming execution with bounded Rayon flush chunks of size $B$ requires $O(B + |\mathcal{S}|)$ working memory independent of total row count $N = \sum_E N_E$, while guaranteeing $100\%$ foreign-key referential integrity ($f_{\text{PK}}(s, P, i_P) \in \{f_{\text{PK}}(s, P, j) : 0 \le j < N_P\}$) and $100\%$ satisfaction of all multi-hop \code{@after} temporal constraints.
\end{proposition}
\begin{proof}
By Eq.~\eqref{eq:fksample}, every sampled parent index satisfies $0 \le i_P < N_P$. Since parent entity $P$ emits exactly the set of primary keys $\{f_{\text{PK}}(s, P, j) : 0 \le j < N_P\}$, every child foreign key $f_{\text{PK}}(s, P, i_P)$ is guaranteed to exist in $P$'s output table. Similarly, because $\text{evalDate}(s, P, i_P, \text{parentDate})$ is a deterministic function of $(s, P, i_P, \text{parentDate})$, re-evaluating it inside child $C$ yields the exact epoch day $t_P$ written to row $i_P$ of $P$, so $t_C = t_P + \Delta \in [t_P + d_{\min}, t_P + d_{\max}]$ holds identically. Each recursive call allocates only $O(d)$ stack frames, and worker threads flush batches of $B = 16{,}384$ rows to buffered disk writers with on-the-fly SHA-256 hashing, bounding resident memory by $O(B + |\mathcal{S}|)$.
\end{proof}

\section{Empirical Evaluation}\label{sec:eval}

We evaluate Samyama-SDG (\code{sdg-rs} v0.2.0, compiled with \code{rustc --release}, LTO enabled, 16 Rayon threads) on a Linux workstation to answer five questions:
\begin{enumerate}
\item \textbf{Q1 (Throughput \& $O(1)$ Memory Scaling):} How do generation latency, throughput, and peak resident set size (RSS, measured via Linux kernel high-water mark \code{VmHWM}) scale from $1{,}000$ to $5{,}000{,}000$ records across schemas and output formats?
\item \textbf{Q2 (Referential, Regex, and Temporal Conformance):} Does the $O(1)$-memory index-bijective engine achieve $0$ conformance errors across millions of foreign keys, regex patterns, and 4-hop temporal chains?
\item \textbf{Q3 (Foreign-Key Degree Realism):} How effectively does \code{@fk(dist: "zipf", s)} model power-law hub concentration compared to uniform foreign keys?
\item \textbf{Q4 (Layer 2 Causal Pack Acceleration):} What speedup does native Rust execution provide for the Layer 2 Telecom NOC causal alarm-storm generator compared to the Python reference implementation?
\item \textbf{Q5 (Fit--Generate--Audit):} Starting from a real table, can a differentially private fit followed by forward compilation produce data that is useful for a downstream task, while an empirical audit finds no sign of record copying?
\end{enumerate}

\begin{table*}[t]
\centering
\footnotesize
\setlength{\tabcolsep}{4.5pt}
\caption{Multi-Scale Generation Latency, Throughput, and Peak Resident Memory (\code{VmHWM}) Across Three Schemas (\code{seed = 20261019})}
\label{tab:scaling}
\makebox[\textwidth][c]{%
\begin{tabular}{@{}llrrrrrr@{}}
\toprule
\textbf{Schema} & \textbf{Output Mode} & \textbf{Scale} & \textbf{Total Rows} & \textbf{Graph Edges} & \textbf{Wall Time (ms)} & \textbf{Throughput (M rows/s)} & \textbf{Peak RSS (MiB)} \\
\midrule
Health Clinic ($5$ tables) & CSV & $1\times$ & $1{,}000$ & $0$ & $2.15$ & $0.47$ & $7.7$ \\
Health Clinic ($5$ tables) & CSV & $10\times$ & $10{,}000$ & $0$ & $5.95$ & $1.68$ & $7.9$ \\
Health Clinic ($5$ tables) & CSV & $100\times$ & $100{,}000$ & $0$ & $27.95$ & $3.58$ & $18.0$ \\
Health Clinic ($5$ tables) & CSV & $1{,}000\times$ & $1{,}000{,}000$ & $0$ & $\mathbf{172.43}$ & $\mathbf{5.80}$ & $\mathbf{35.3}$ \\
Health Clinic ($5$ tables) & CSV & $5{,}000\times$ & $5{,}000{,}000$ & $0$ & $\mathbf{808.56}$ & $\mathbf{6.18}$ & $\mathbf{35.1}$ \\
\midrule
Health Clinic ($5$ tables) & CSV + Graph JSONL & $1\times$ & $1{,}000$ & $1{,}150$ & $2.90$ & $0.34$ & $8.1$ \\
Health Clinic ($5$ tables) & CSV + Graph JSONL & $10\times$ & $10{,}000$ & $11{,}500$ & $13.99$ & $0.71$ & $12.2$ \\
Health Clinic ($5$ tables) & CSV + Graph JSONL & $100\times$ & $100{,}000$ & $115{,}000$ & $63.52$ & $1.57$ & $46.2$ \\
Health Clinic ($5$ tables) & CSV + Graph JSONL & $1{,}000\times$ & $1{,}000{,}000$ & $1{,}150{,}000$ & $\mathbf{438.70}$ & $\mathbf{2.28}$ & $125.7$ \\
Health Clinic ($5$ tables) & CSV + Graph JSONL & $5{,}000\times$ & $5{,}000{,}000$ & $5{,}750{,}000$ & $\mathbf{1{,}944.36}$ & $\mathbf{2.57}$ & $132.3$ \\
\midrule
E-Commerce ($4$ tables) & CSV & $1{,}000\times$ & $1{,}000{,}000$ & $0$ & $\mathbf{128.79}$ & $\mathbf{7.76}$ & $\mathbf{31.4}$ \\
E-Commerce ($4$ tables) & CSV & $5{,}000\times$ & $5{,}000{,}000$ & $0$ & $\mathbf{676.12}$ & $\mathbf{7.40}$ & $\mathbf{31.4}$ \\
\midrule
University ($4$ tables) & CSV & $1{,}000\times$ & $1{,}000{,}000$ & $0$ & $\mathbf{147.15}$ & $\mathbf{6.80}$ & $\mathbf{31.1}$ \\
University ($4$ tables) & CSV & $5{,}000\times$ & $5{,}000{,}000$ & $0$ & $\mathbf{698.24}$ & $\mathbf{7.16}$ & $\mathbf{32.4}$ \\
\bottomrule
\end{tabular}}
\end{table*}

\begin{figure*}[t]
\centering
\includegraphics[width=0.96\textwidth]{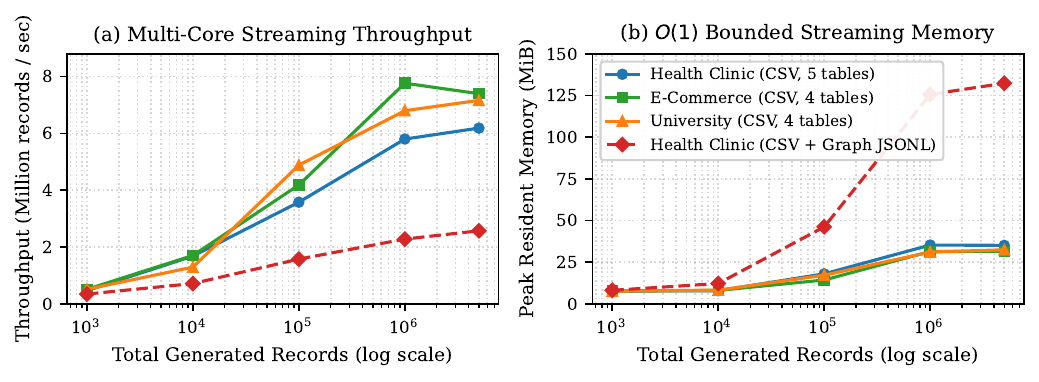}
\caption{Multi-scale throughput (a) and peak resident memory \code{VmHWM} (b) from $1{,}000$ to $5{,}000{,}000$ records. Between $1\text{M}$ and $5\text{M}$ records, CSV resident memory stays completely flat ($31.4\text{ MiB}$--$35.1\text{ MiB}$), empirically confirming Proposition~\ref{prop:o1fk}.}
\label{fig:scaling}
\end{figure*}

\subsection{Throughput and $O(1)$ Memory Scaling (Q1)}

We benchmark three multi-table GraphQL schemas:
\begin{itemize}
\item \textbf{Health Clinic (\code{health\_clinic.graphql}):} $5$ tables (\code{Patient}, \code{Doctor}, \code{Appointment}, \code{Diagnosis}, \code{Claim}), $6$ foreign-key edges, $4$-hop \code{@after} temporal chain (\code{dob} $\to$ \code{appointmentDate} $\to$ \code{recordedDate} $\to$ \code{filedDate}), and clinical regexes (\code{SSN}, \code{NPI}, \code{ICD10}, \code{CPT}).
\item \textbf{E-Commerce (\code{ecommerce.graphql}):} $4$ tables (\code{Customer}, \code{Product}, \code{Order}, \code{OrderItem}), Zipfian customer-order and product-item skew, \code{SKU} and \code{IPv4} regexes, and log-normal prices.
\item \textbf{University (\code{university.graphql}):} $4$ tables (\code{Department}, \code{Course}, \code{Student}, \code{Enrollment}) with ULID and prefixed primary keys and Zipfian course enrollment.
\end{itemize}

Table~\ref{tab:scaling} and Fig.~\ref{fig:scaling} report wall-clock time, throughput, and peak RSS across scales $1\times$ ($1{,}000$ rows) to $5{,}000\times$ ($5{,}000{,}000$ rows). In CSV mode, \code{sdg-rs} generates $1{,}000{,}000$ records in $128.79\text{ ms}$ ($7.76\text{M}$ rows/s) on E-Commerce, $147.15\text{ ms}$ ($6.80\text{M}$ rows/s) on University, and $172.43\text{ ms}$ ($5.80\text{M}$ rows/s) on Health Clinic. At $5{,}000{,}000$ records, Health Clinic completes in $808.56\text{ ms}$ ($6.18\text{M}$ rows/s). Crucially, peak resident memory between $1{,}000{,}000$ rows and $5{,}000{,}000$ rows is invariant: $35.3\text{ MiB}$ vs.\ $35.1\text{ MiB}$ on Health Clinic and $31.4\text{ MiB}$ vs.\ $31.4\text{ MiB}$ on E-Commerce. In dual CSV + Property Graph JSONL mode (which emits $5{,}000{,}000$ CSV rows, $5{,}000{,}000$ JSONL node objects, and $5{,}750{,}000$ JSONL edge objects with on-the-fly SHA-256 digests), $1{,}000{,}000$ records complete in $438.70\text{ ms}$ ($2.28\text{M}$ rows/s) and $5{,}000{,}000$ records complete in $1{,}944.36\text{ ms}$ ($2.57\text{M}$ rows/s) at $125.7\text{ MiB}$--$132.3\text{ MiB}$ RSS.

\subsection{Conformance and Foreign-Key Degree Realism (Q2, Q3)}

Running \code{sdg-rs validate-schema} on the $1{,}000{,}000$-row Health Clinic dataset verifies all $1{,}000{,}000$ primary keys, $1{,}150{,}000$ foreign-key references, $450{,}000$ regex-constrained attributes, and $900{,}000$ multi-hop \code{@after} date constraints, reporting \textbf{$0$ conformance errors}.

\begin{figure}[t]
\centering
\includegraphics[width=0.95\columnwidth]{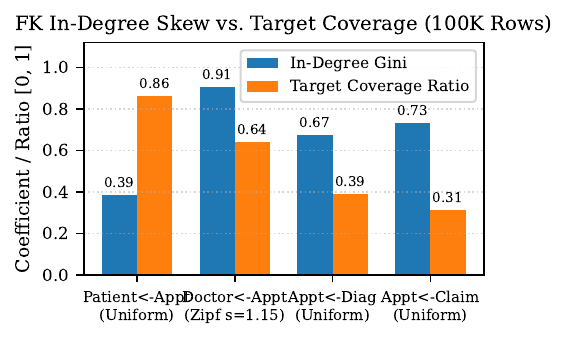}
\caption{Foreign-key in-degree Gini coefficient and target entity coverage ratio on the $100{,}000$-row Health Clinic dataset.}
\label{fig:fk}
\end{figure}

Table~\ref{tab:fk} and Fig.~\ref{fig:fk} compare the in-degree distributions of uniform vs.\ Zipfian foreign keys on the $100{,}000$-row Health Clinic dataset. While \code{Appointment.patient -> Patient} (\code{uniform}) yields an even degree distribution (Gini $0.386$, max in-degree $10$, $0.865$ target coverage), \code{Appointment.doctor -> Doctor} (\code{zipf}, $s = 1.15$) produces realistic specialist hub concentration (Gini $0.907$, max in-degree $8{,}173$, $0.642$ target coverage), matching power-law degree skews observed in clinical referral graphs without sacrificing $O(1)$-memory generation.

\begin{table}[t]
\centering
\footnotesize
\setlength{\tabcolsep}{2.6pt}
\caption{Foreign-Key In-Degree Distribution Realism on Health Clinic ($100{,}000$ Rows, $115{,}000$ Edges)}
\label{tab:fk}
\begin{tabular}{@{}llrrrr@{}}
\toprule
\textbf{Foreign Key Edge} & \textbf{Dist.} & \textbf{Coverage} & \textbf{p90} & \textbf{Max} & \textbf{Gini} \\
\midrule
\code{Appt.patient -> Patient} & Uniform & $0.865$ & $4$ & $10$ & $0.386$ \\
\code{Appt.doctor -> Doctor} & Zipf $1.15$ & $0.642$ & $6$ & $8{,}173$ & $\mathbf{0.907}$ \\
\code{Diag.appt -> Appt} & Uniform & $0.393$ & $1$ & $5$ & $0.674$ \\
\code{Claim.appt -> Appt} & Uniform & $0.312$ & $1$ & $6$ & $0.733$ \\
\bottomrule
\end{tabular}
\end{table}

\subsection{Layer 2 Telecom NOC Causal Generator: Rust vs.\ Python (Q4)}

Table~\ref{tab:telecom} compares the native Rust Layer 2 Telecom NOC causal alarm-storm generator (\code{src/telecom.rs}) against the Python reference implementation on G\'{E}ANT 2012 (\code{small.yaml}: $70$ faults, $40$ scored eval incidents; and \code{large.yaml}: $350$ faults, $200$ scored eval incidents). Both implementations bind $100\%$ of node labels and alarm enumerations to \code{samyama-ontologies} with $0$ conformance errors, while the Rust implementation reduces end-to-end topology expansion, fault propagation, alarm-storm synthesis, and JSONL serialization time from $332.44\text{ ms}$ to $124.70\text{ ms}$ ($2.67\times$ speedup) on \code{small.yaml} and from $854.92\text{ ms}$ to $429.98\text{ ms}$ ($1.99\times$ speedup) on \code{large.yaml}.

\begin{table}[t]
\centering
\footnotesize
\setlength{\tabcolsep}{2.8pt}
\caption{Layer 2 Telecom NOC Causal Pack Execution Time: Rust (\code{sdg-rs}) vs.\ Python (\code{sdg})}
\label{tab:telecom}
\makebox[\columnwidth][c]{%
\begin{tabular}{@{}lrrrrr@{}}
\toprule
\textbf{Config} & \textbf{Faults} & \textbf{Eval Inc.} & \textbf{Python (ms)} & \textbf{Rust (ms)} & \textbf{Speedup} \\
\midrule
\code{small.yaml} & $70$ & $40$ & $332.44$ & $\mathbf{124.70}$ & $\mathbf{2.67\times}$ \\
\code{large.yaml} & $350$ & $200$ & $854.92$ & $\mathbf{429.98}$ & $\mathbf{1.99\times}$ \\
\bottomrule
\end{tabular}}
\end{table}

\subsection{End-to-End Fit, Generate, and Audit on a Public Table (Q5)}\label{sec:e2e}

Sections~\ref{sec:lang}--\ref{sec:fk} generate data from a specification a person wrote. To test the pipeline that starts from data, we fit UCI Adult~\cite{becker1996adult} ($32{,}561$ training rows, CC~BY~4.0), binned with fixed public cut points into $13$ categorical columns, and generate from the fitted specification. The fit adds Gaussian noise to one-way marginals and to the pairwise marginals of every column with the income label under $\rho$-zCDP, converted to $(\epsilon,\delta)=(1,10^{-6})$, with one row per entity; its structure (a tree rooted at the label) and category domains are fixed in advance, so no budget is spent on structure. The result is a \code{.graphql} file holding only noisy parameters; it, the evaluation harness (\code{sdg-eval}) and the script that reproduces Table~\ref{tab:e2e} are in the public repository at tag \code{v2.1.0} (\code{examples/adult\_e2e}), while the fitting step itself is not released. The specification is compiled by \code{sdg-rs} with the \code{@given(field, table)} directive, which samples a column conditional on another column of the same row. The parent value is recomputed from $(\text{seed},\text{entity},\text{row},\text{field})$, so the $O(1)$-memory property of Section~\ref{sec:fk} is unchanged.

We compare three sets of $32{,}561$ rows against $16{,}281$ real rows the fitter never saw: the generated data, a \emph{leaky} control (a bootstrap of the training rows, which the audit must flag) and a \emph{naive} control (independent resampling of the one-way marginals, which the utility test must penalise). Pass criteria were fixed before the first run. Utility is the AUC of a random forest trained on a set and tested on the unseen real rows (TSTR); the real-trained ceiling (TRTR) is $0.908$. The audit scores each candidate record by its Hamming distance to the nearest generated row and reports the AUC of separating $16{,}281$ training rows from the $16{,}281$ unseen rows.

\begin{table}[t]
\centering
\footnotesize
\setlength{\tabcolsep}{3.2pt}
\caption{Fit--generate--audit on UCI Adult at $(\epsilon,\delta)=(1,10^{-6})$. TSTR: AUC when trained on the set and tested on unseen real rows. Gap: mean pairwise-association difference from the real data. MIA: membership-inference AUC ($0.5$ means no signal). Exact: share of rows identical to a training row; fresh real rows score $0.495$.}
\label{tab:e2e}
\begin{tabular}{@{}lrrrr@{}}
\toprule
\textbf{Set} & \textbf{TSTR} & \textbf{Gap} & \textbf{MIA} & \textbf{Exact} \\
\midrule
Generated (DP fit) & $\mathbf{0.882}$ & $0.076$ & $\mathbf{0.503}$ & $0.113$ \\
Leaky (bootstrap) & $0.907$ & $0.003$ & $0.680$ & $1.000$ \\
Naive (independent) & $0.524$ & $0.142$ & $0.501$ & $0.065$ \\
\bottomrule
\end{tabular}
\end{table}

Table~\ref{tab:e2e} shows that the generated data reaches $0.882$ against the $0.908$ ceiling (ratio $0.97$), while independent columns reach $0.524$: the pairwise structure is what carries the task. The audit separates members from non-members for the leaky control ($0.680$) but not for the generated data ($0.503$), which is also farther from the training rows (mean distance $1.48$) than fresh real rows are ($0.64$). Three limits apply. Dependencies with the label are kept (mean gap $0.006$), but dependencies between other columns are only partly kept: the mean pairwise gap is $0.076$ and the worst pair differs by $0.556$, because the tree models each column given the label alone. The table is a single relation, so foreign-key structure is not exercised. And the audit detects copying, not every leakage channel; the guarantee is the $(\epsilon,\delta)$ ledger, and the audit is a check on top of it.

\section{Conclusion}\label{sec:conclusion}

Samyama-SDG bridges declarative schema-driven data generation and ontology-bound causal simulation in a single compiled Rust engine. By combining Cardinality-Extended GraphQL SDL, a global Attribute Data Dictionary, stateless counter-based seekable PRNGs, and $O(1)$-memory index-bijective foreign keys, \code{sdg-rs} generates multi-million-row relational and property-graph datasets in sub-second wall time and flat resident memory with $100\%$ referential, regex, ontology, and temporal conformance.

\paragraph{Composition with Inverse Database Profiling and Privacy Scope.}
While \code{sdg-rs} is a forward compiler from explicit schemas and distribution parameters to synthetic tables and graphs, its declarative parameterization composes orthogonally with inverse database fitting. Specifically, dataless catalog and plan-cloning regenerators such as Hydra~\cite{sanghi2018hydra} as well as $(\epsilon, \delta)$-differentially private marginal mechanisms---including MST~\cite{mckenna2021mst}, AIM~\cite{mckenna2022aim}, and multi-relational PrivLava~\cite{cai2023privlava}---can emit sanitized schema cardinality, skew, and attribute-dictionary parameters $\hat{\Theta}$ from private production databases. Because forward compilation and streaming in \code{sdg-rs} execute deterministically over $(\hat{\Theta}, s)$ without touching raw tuples, materializing datasets of arbitrary scale incurs zero additional privacy loss ($\Delta\epsilon = 0$) by the differential privacy post-processing theorem~\cite{dwork2014algorithmic}. Consistent with Stadler et al.~\cite{stadler2022groundhog} and the computational hardness bounds of Ullman and Vadhan~\cite{ullman2011pcps}, forward synthetic generation provides formal privacy guarantees only when composed with explicit $(\epsilon, \delta)$-DP noise calibration over bounded-sensitivity low-order marginals; Section~\ref{sec:e2e} evaluates one such composition end to end on a single public table; multi-table fitting and higher-order dependencies are left to future work.

\bibliographystyle{IEEEtran}
\bibliography{refs}

\end{document}